# A Comparative Study Between Silicon Carbide and Silicon Nitride Based Single Cell CMUT

**Rakesh Kanjilal[1]****, Reshmi Maity[2]**

Electronics and Communication Engineering, Mizoram University, Aizawl, India
**E-mail:** [1]mailtokanji@gmail.com, [2] reshmidas_2009@rediffmail.com

**Abstract**

This paper explores the design and conducts a comparative analysis of a non-insulated Capacitive Micromachined Ultrasonic Transducer (CMUT) featuring an innovative asymmetric electrode configuration to improve the performance of the device. Specifically, this configuration involves the utilization of a top electrode with a smaller radius in comparison to the bottom electrode. The study encompasses an investigation into the effects of varying biasing voltage within the range of 40 V to 100 V. The materials employed in this study are carefully selected to optimize the CMUT's performance. The substrate material is silicon, and the bottom and top electrodes are made from aluminium. Additionally, silicon dioxide is utilized as the pillar material within the device's structure.

**Keywords:** CMUT - Principle of Operation, Silicon Carbide & Silicon Nitride based membrane, Displacement, Frequency, Capacitance, COMSOL Multiphysics.

1. **Introduction**

CMUT ( Capacitive Micromachined Ultrasonic Transducer) is an ultrasonic transducer, The basic working principle is the generation and reception of ultrasonic waves using the principle of capacitance and mechanical deformation. MEMS research began in the 1960s, focusing on miniaturized mechanical and electronic devices, The concept of CMUTs was first introduced in the early 1990s by researchers at Stanford University. From the start of the 2000s to the Present CMUT gained significant attention in the field of medical imaging, particularly in the development of high–frequency, higher resolution ultrasound probes. From the 2010s to the Present researchers also working on integrating CMUTs with Complimentary metal-oxide-semiconductor (CMOS) electronics, allowing for on-chip signal processing and miniaturization of ultrasound systems with a focus on further improvement of performance, reducing the cost and exploring new applications.

## 2. Analytical Model

The calculation-based capacitance analysis that has been done totally excluded the fringing effect of the device, for analysis of the capacitance and for the whole device equivalent capacitance substitute them according to the orientation of layers, The basic capacitance formula used for the calculation:

$$C = K\frac{\epsilon A}{d} \tag{1}$$

Where, C is the Capacitance, K is the dielectric constant of free space, $\epsilon$ is the Relative permittivity of the material, A is the area and d is the distance between the top and bottom electrode.

For calculating Equivalent device capacitance,

$$C_{eq} = (C_a \times C_g) / (C_a + C_g) \tag{2}$$

Where, $C_a$ is the capacitance of the actuation layer and $C_g$ is the capacitance of the gap.

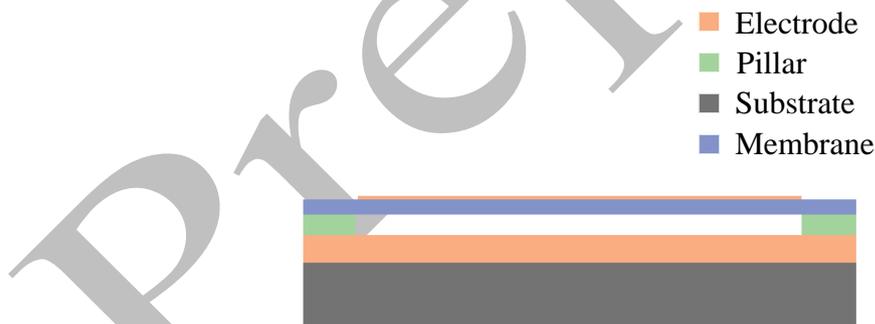

**Figure 1.** Structure of Non-Insulated single cell CMUT

At the heart of a CMUT lies a flexible actuation layer that plays a crucial role in the transducer's operation. The displacement of this actuation layer is a fundamental aspect of CMUT functionality, as it is responsible for generating and detecting ultrasonic waves. The displacement of the actuation layer in a CMUT is induced by an applied voltage. An electrostatic force is generated when a voltage is applied between the membrane and an underlying electrode. This force causes the actuation layer to deform or move, displacing the membrane away from the electrode.

According to Mason's analysis of membrane deflection of non-insulated structure, the expression under tension will be,

$$\frac{(Y_a+T_a)t_a^3}{12\,(1-\sigma_a^2)}\nabla^4 w_m - T_a\nabla^2 w_m - P + t_a\rho_a\frac{\partial^2 w_m}{\partial t^2} = 0 \qquad (3)$$

Here $t_a$ is the thickness of actuation layer, $w_m$ is the displacement of the actuation layer, $\rho_a$ is density of actuation material, $T_a$ is the Tension.

### 3. FEM Model

Using COMSOL Multiphysics, constructed a sophisticated 3D simulation model for the study. This enabled to delve deep into the intricacies of the problem, ensuring meaningful results.

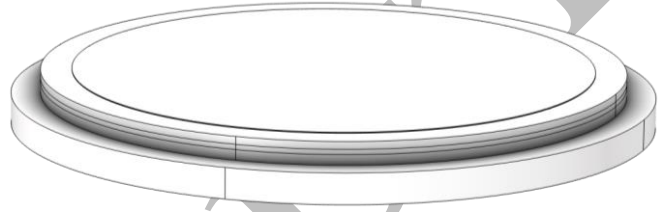

**Figure 2.** Simulated 3D single cell CMUT.

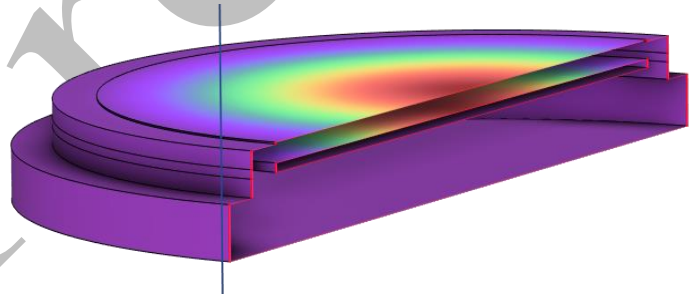

**Figure 3.** Cross-sectional view of the simulated 3D cell.

For the designing purpose we took some material properties for designing the structure and also listed all the parameters that has been used for simulation in the Table 1 & Table 2 respectively.

**Table 1.** Properties of the materials used for the simulation

| Materials | Properties | | | |
|---|---|---|---|---|
| | *Density (kg/m)* | *Relative Permitivity* | *Young's Modulus (Gpa)* | *Poisson's ratio* |
| Silicon (Si) | 2329 | 11.7 | 170 | 0.28 |
| Silicon Nitride ($Si_3N_4$) | 3100 | 7.5 | 250 | 0.23 |
| Silicon Carbide (SiC) | 3216 | 9.7 | 748 | 0.45 |
| Silicon Dioxide ($SiO_2$) | 2200 | 3.9 | 70 | 0.17 |
| Aluminium (Al) | 2700 | - | 70 | 0.33 |

**Table 2.** Parameters of the Single cell CMUT Model

| Parameters | Details | |
|---|---|---|
| | *Value (μm)* | *Description* |
| subR | 28 | Substrate Radius |
| subH | 3 | Substrate Thickness |
| belecH | 1 | Bottom Electrode Thickness |
| oxiH | 0.5 | Airgap Height |
| oxiinR | 25 | Inner Cavity Radius |
| telecH | 0.1 | Top Electrode Thickness |
| memH | 0.75 | Membrane Thickness |

## 4. Results and Discussion
### 4.1. Capacitance Analysis

When the cavity radius increases in the non-insulated CMUT design with a vertically placed oxide pillar, the gap between the top and bottom electrodes remains relatively constant or decrease slightly. The decrease in capacitance is primarily caused by the reduction in overlap area between the electrodes due to the increase in oxide pillar radius. The plot of the capacitance decreasing behaviour is shown below at Figure 4 and Figure 5 for silicon nitride and silicon carbide material-based actuation layer respectively and the comparative result also shown in Table 3.

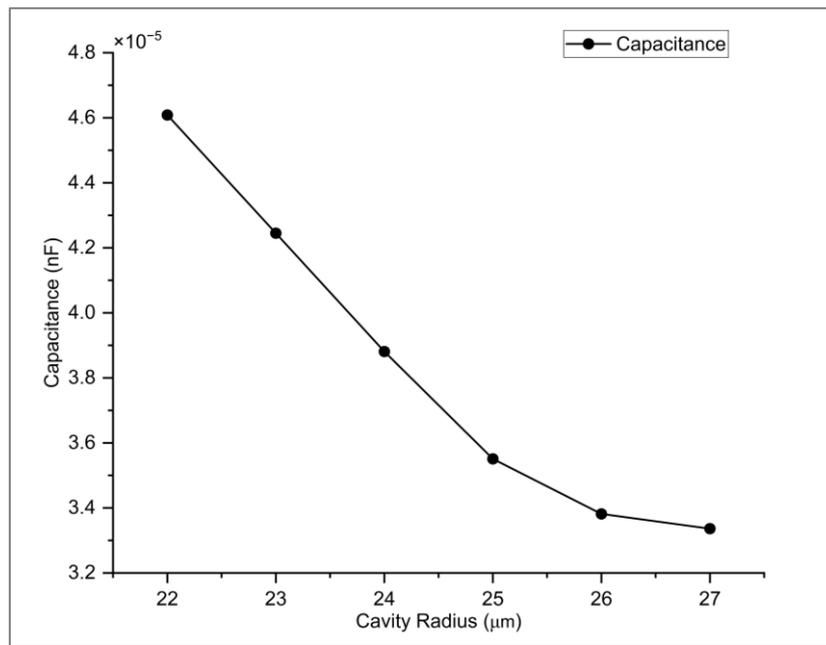

**Figure 4.** Capacitance Vs Cavity Radius plot of $Si_3N_4$ membrane-based cell

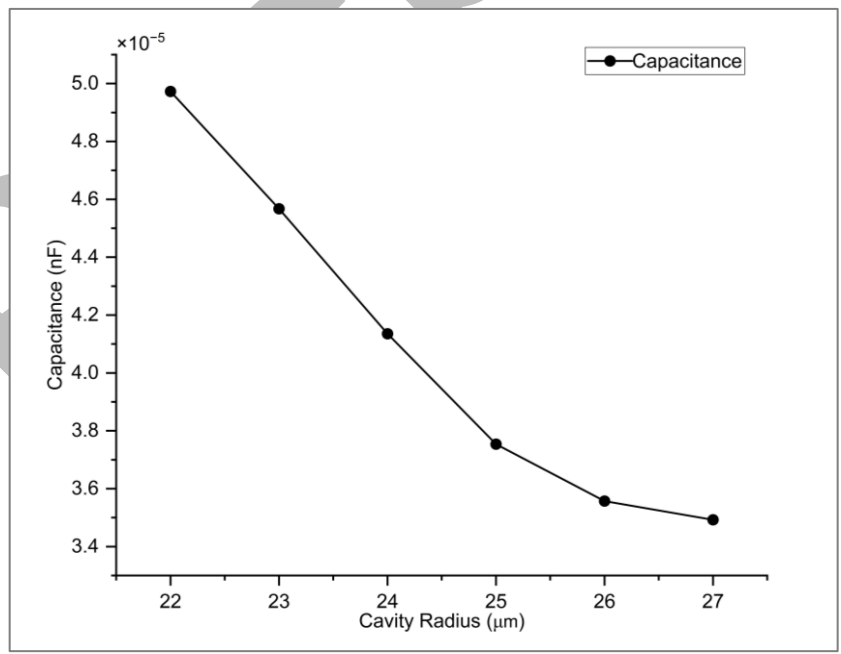

**Figure 5.** Capacitance Vs Cavity Radius plot of SiC membrane-based cell

**Table 3.** Capacitance data of $Si_3N_4$ & SiC based cell by varying Cavity Radius

| Cavity Radius (μm) | *Capacitance(nF)* *($Si_3N_4$)* | *Capacitance(nF)* *(SiC)* |
|---|---|---|
| 22 | $4.6086 \times 10^{-5}$ | $4.9729 \times 10^{-5}$ |
| 23 | $4.2449 \times 10^{-5}$ | $4.5670 \times 10^{-5}$ |
| 24 | $3.8811 \times 10^{-5}$ | $4.1353 \times 10^{-5}$ |
| 25 | $3.5508 \times 10^{-5}$ | $3.7537 \times 10^{-5}$ |
| 26 | $3.3814 \times 10^{-5}$ | $3.5572 \times 10^{-5}$ |
| 27 | $3.3362 \times 10^{-5}$ | $3.4925 \times 10^{-5}$ |

The decrease in capacitance with increasing cavity thickness in the non-insulated CMUT is attributed to the increased distance between the electrodes and the presence of the oxide pillar, which reduces the electric field strength and affects the effective capacitance of the device, in the Figure 6 and the Table 4. the plot of the capacitance and respective values demonstrated accordingly.

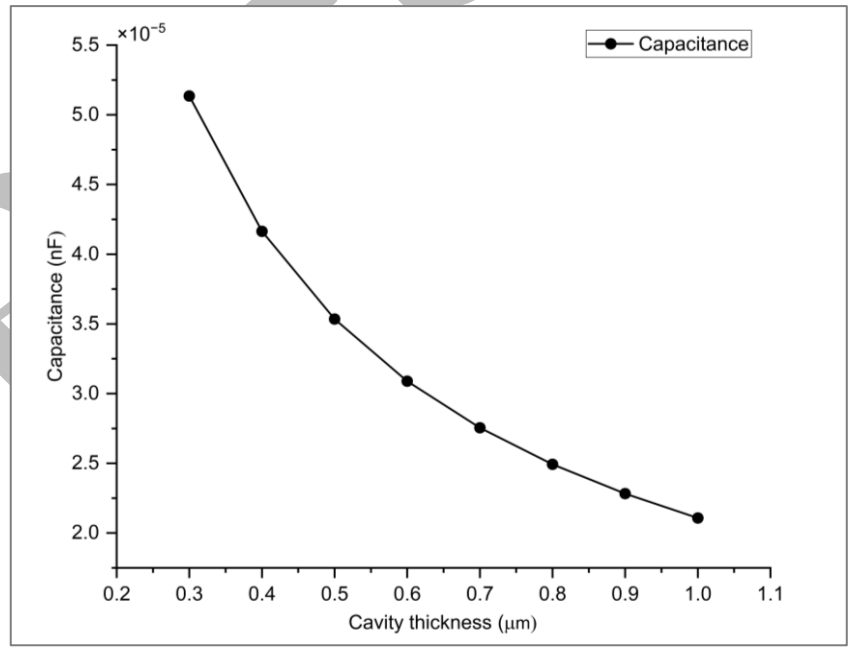

**Figure 6.** Capacitance Vs Cavity thickness plot of $Si_3N_4$ membrane-based cell

**Table 4.** Capacitance data of $Si_3N_4$ & SiC based cell

by varying Cavity thickness

| Cavity Thickness (µm) | *Capacitance(nF)* *($Si_3N_4$)* | *Capacitance(nF)* *(SiC)* |
|---|---|---|
| 0.3 | $5.1348 \times 10^{-5}$ | $5.4493 \times 10^{-5}$ |
| 0.4 | $4.1641 \times 10^{-5}$ | $4.4180 \times 10^{-5}$ |
| 0.5 | $3.5353 \times 10^{-5}$ | $3.7537 \times 10^{-5}$ |
| 0.6 | $3.0894 \times 10^{-5}$ | $3.2759 \times 10^{-5}$ |
| 0.7 | $2.7548 \times 10^{-5}$ | $2.9157 \times 10^{-5}$ |
| 0.8 | $2.4931 \times 10^{-5}$ | $2.6346 \times 10^{-5}$ |
| 0.9 | $2.2825 \times 10^{-5}$ | $2.4085 \times 10^{-5}$ |
| 1 | $2.1081 \times 10^{-5}$ | $2.2222 \times 10^{-5}$ |

4.**2. Frequency Analysis**

the eigenfrequencies for the first four modes of the CMUT model provide valuable insights into its structural behaviour.

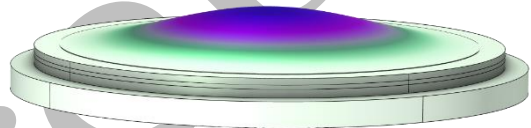

**Figure 7.** Eigenfrequency of the device 1st mode

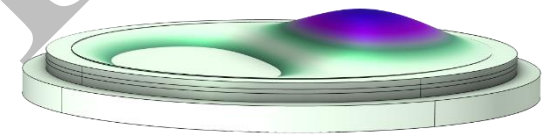

**Figure 8.** Eigenfrequency of the device 2nd mode

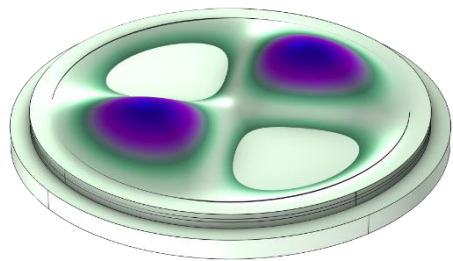

**Figure 9.** Eigenfrequency of the device 3rd mode

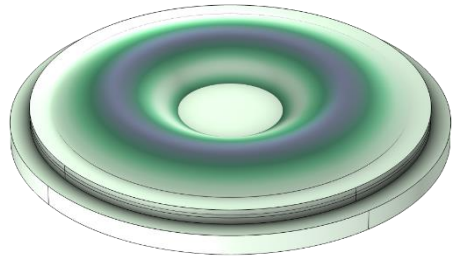

**Figure 10.** Eigenfrequency of the device 4th mode

Eigenfrequencies, also known as natural frequencies or resonant frequencies, are important parameters in the analysis of vibrating structures like CMUTs. These frequencies represent the inherent vibrational modes of the structure. Calculating eigenfrequencies for a CMUT model involves solving a structural mechanics problem, and the specific factors and methods can vary depending on the complexity of the model. Eigenfrequencies are typically determined using numerical methods like finite element analysis (FEA). FEA involves discretizing the CMUT structure into a mesh of smaller elements. The more refined the mesh, the more accurate the results. The CMUT structure is then modeled as a set of interconnected elements with defined material properties and boundary conditions.

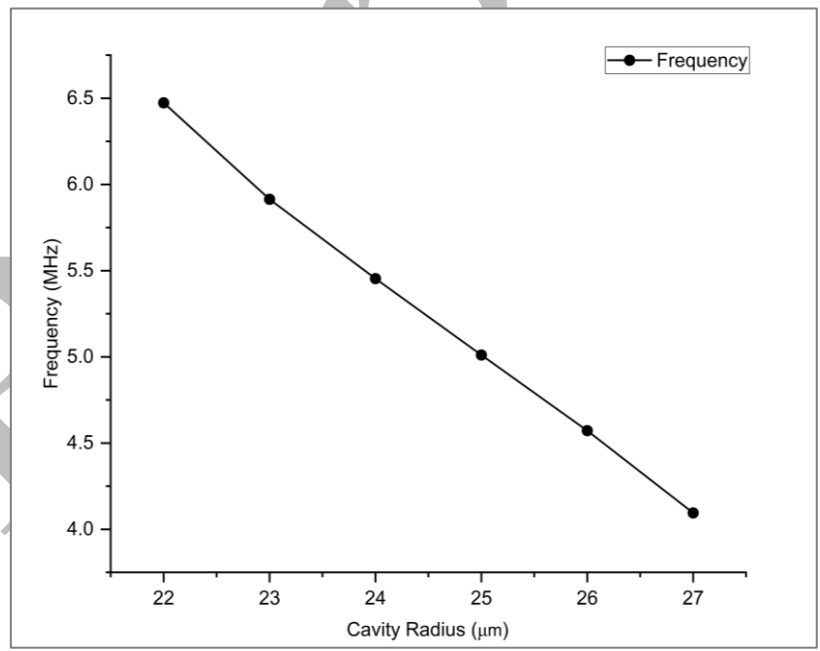

**Figure 11.** Frequency Vs Cavity Radius of $Si_3N_4$ membrane based CMUT

The overall separation between the top and bottom electrodes not significantly changes with the increase in cavity radius. Therefore, the decrease in capacitance observed primarily due to the reduction in overlap area between the electrodes caused by the increase in oxide pillar radius.

The comparative frequency (1st mode) data is demonstrated in the Table 5.

**Table 5.** Frequency data of $Si_3N_4$ & SiC based cell varying cavity radius

| Cavity Radius ($\mu m$) | Frequency (MHz)) ($Si_3N_4$) | Frequency (MHz) (SiC) |
|---|---|---|
| 22 | 6.4731 | 10.888 |
| 23 | 5.9142 | 10.004 |
| 24 | 5.4535 | 9.2114 |
| 25 | 5.0105 | 8.4814 |
| 26 | 4.5719 | 7.7328 |
| 27 | 4.0949 | 6.6516 |

The increase in resonant frequency with an increase in membrane thickness in the non-insulated CMUT is due to the increased stiffness introduced by the thicker membrane.

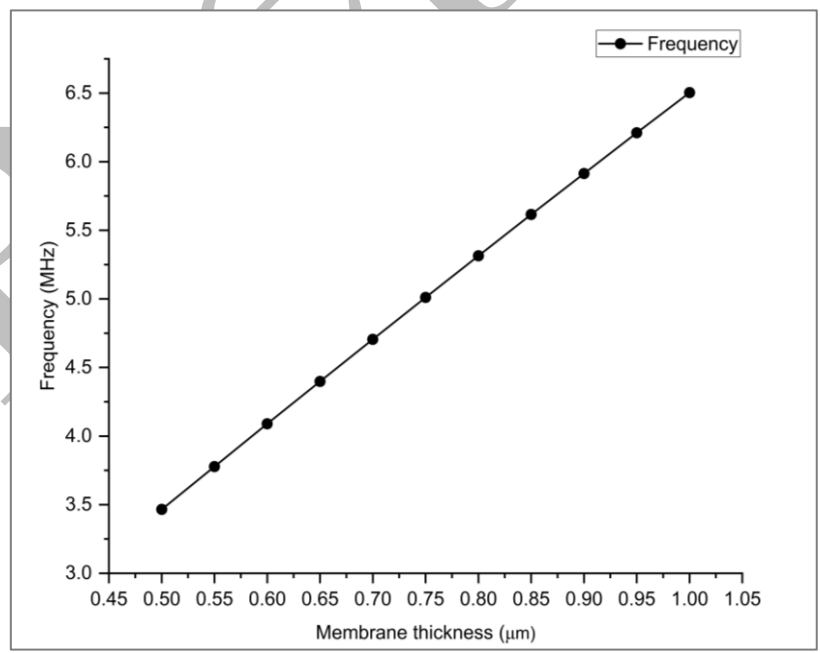

**Figure 12.** Frequency Vs Membrane thickness of $Si_3N_4$ membrane based CMUT cell.

When the cavity thickness of the non-insulated CMUT increases, the frequency response shifts towards higher frequencies due to the decrease in capacitance and the corresponding increase in the effective stiffness of the device.

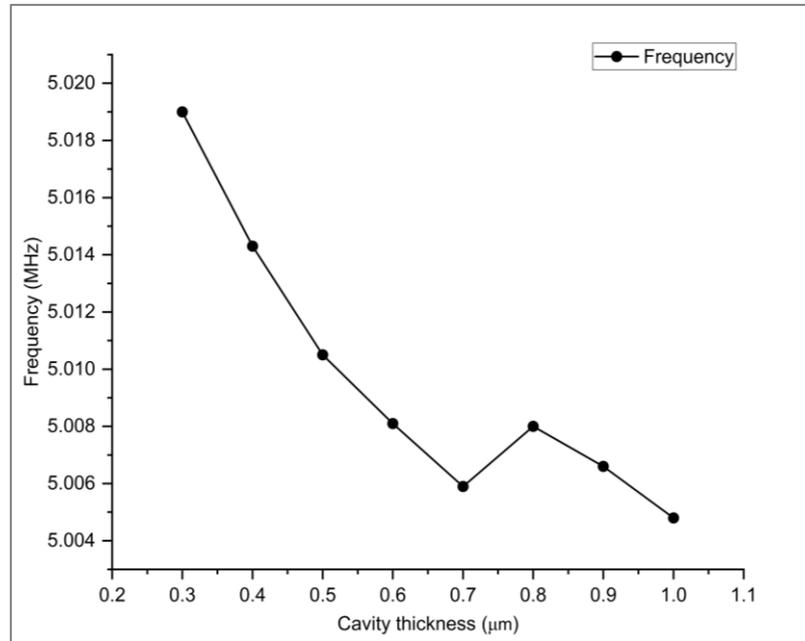

**Figure 13.** Frequency Vs Cavity thickness of $Si_3N_4$ membrane based CMUT cell.

### 4.3. Displacement Analysis

As the applied voltage increases, the electric field strength between the plates or across the CMUT structure also increases. This stronger electric field leads to a greater force acting on the charges. According to Coulomb's law, the force between charges is directly proportional to the electric field strength. Therefore, a higher applied voltage results in a larger force on the charges, causing greater displacement. Figure 14 represents the plot of the displacement by varying different voltage.

**Table 6.** Displacement data of $Si_3N_4$ based cell by varying the Applied Voltage

| Applied Voltage (V) | *Displacement (μm)* |
|---|---|
| 40 | 0.0125 |
| 50 | 0.0196 |
| 60 | 0.0282 |
| 70 | 0.0383 |
| 80 | 0.0501 |
| 90 | 0.0633 |
| 100 | 0.0782 |

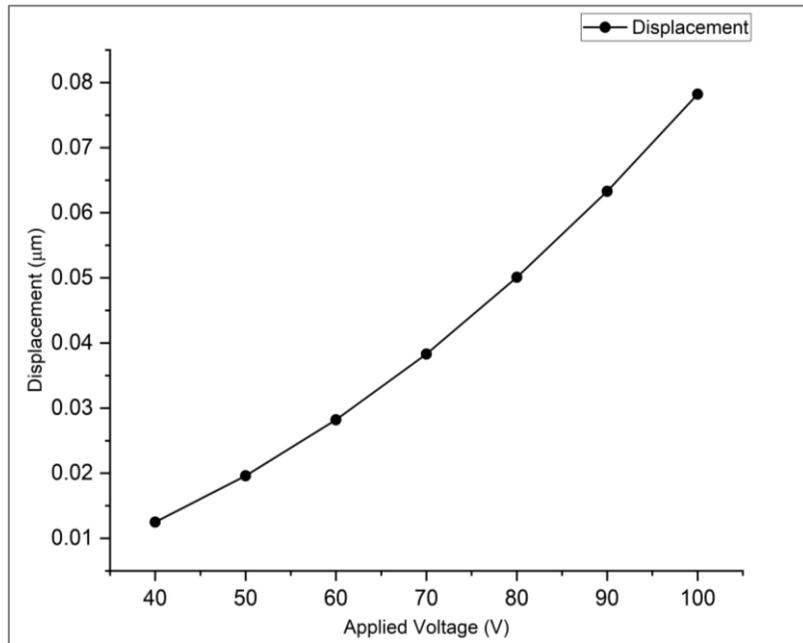

**Figure 14.** Displacement Vs Applied Voltage plot of $Si_3N_4$ membrane-based CMUT cell

In Figure 15 & Figure 16 the displacement of the cell is demonstrated by varying different parameters of actuation layer and Cavity Radius respectively, the applied voltage is 40 V.

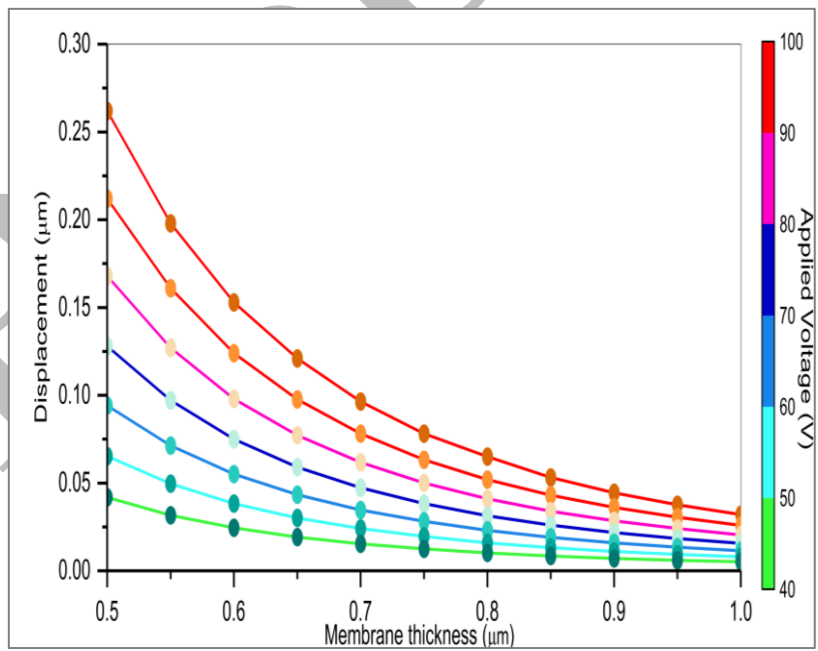

**Figure 15.** Displacement Vs Membrane thickness plot of $Si_3N_4$ membrane-based CMUT cell.

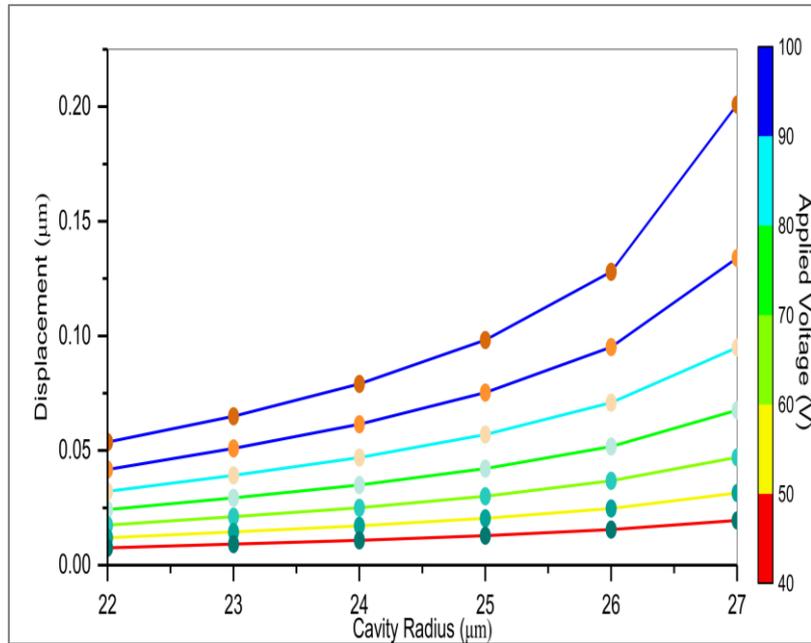

**Figure 16.** Displacement Vs Cavity Radius plot of $Si_3N_4$ membrane-based CMUT cell.

The difference in displacement between $Si_3N_4$ and SiC membrane-based CMUTs at the same applied voltage can be attributed to several material properties and design considerations. In Figure 17 the displacement comparison is demonstrated.

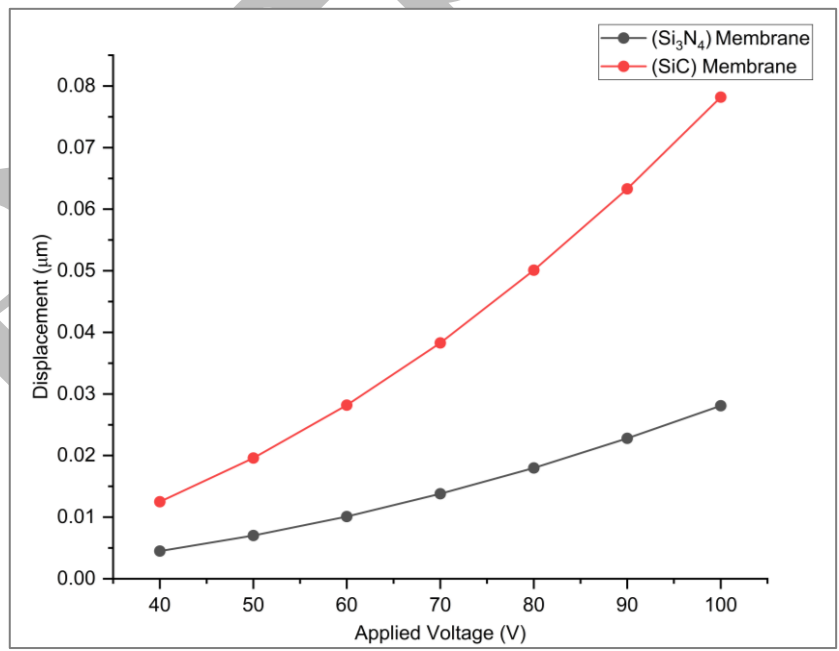

**Figure 17.** Displacement Vs Applied Voltage plot of $Si_3N_4$ membrane-based CMUT cell.

The differences in displacement between $Si_3N_4$ and SiC-based CMUTs at the same applied voltage are likely due to a combination of material properties and design considerations. SiC's greater stiffness and density, along with potential differences in membrane thickness and geometry, could contribute to the observed difference in displacement.

## 5. Conclusion

From the simulated result we have seen the $Si_3N_4$ Based Membrane capacitance and frequency behaviour of the model. According to Figure 4 & Figure 5 that is Capacitance Vs Cavity Radius we have seen capacitance decrease when the cavity radius is increasing, It is the exact opposite behaviour of the capacitance of the device because, in the model top and bottom electrode radius are not same means it is not fully metalized, that's why some overlapping region is there for that the capacitance of the region as known as Parasitic capacitance is also affecting the device capacitance behaviour. From Figure 6, we got the result varying the Cavity thickness and the capacitance behaviour of the device is following the exact expected nature that is when the area of the device is increasing and the distance between the top and bottom electrode is constant the capacitance increases and when the area is constant but distance is increasing then the capacitance of the device decreased. This same result we observed for SiC-based CMUT also. For analyzing the frequency behaviour of the device, we got the expected result that frequency is inversely proportional to capacitance means if the capacitance is increasing frequency decreases, and frequency increases when the capacitance of the model decreases. The plot and values of the frequency behaviour can be observed from Figure 11, Figure 12, Figure 13, and Table 5 accordingly. The same characteristics we have seen for both material $Si_3N_4$ and SiC-based membranes. The behaviour of displacement with respect to the applied voltage in a system depends on various factors, including the specific characteristics of the system and its design. However, in general, the displacement tends to increase with increasing applied voltage. When an electric field is created by applying a voltage to a system, it exerts a force on the charges within the system. The magnitude of this force is directly proportional to the electric field strength. As the applied voltage increases, the electric field strength also increases, resulting in a stronger force acting on the charges. This increased force leads to larger displacements within the system. From Figure 14, we can observe the behaviour and from Figure 15 and Figure 16 we can see the variation of the different layers that is Membrane thickness and cavity radius accordingly of the device but biasing is constant that is 40 V. The same behaviour was observed in both materials.

In every step in the simulation of the CMUT cell, we have seen when comparing the silicon nitride & silicon carbide-based membrane the Capacitance, Frequency, and Displacement are higher in the case of the silicon carbide-based cell compared to silicon nitride. Because SiC has a higher dielectric constant compared to $Si_3N_4$, which means it can store more electric charge per unit area. As a result, SiC membranes tend to have higher capacitance values. The higher capacitance allows for a larger charge storage capacity, enabling better charge accumulation and improved transduction efficiency in CMUT devices.

**Author's biography**

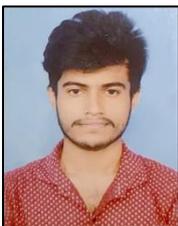

**Rakesh Kanjilal** [1], a recent graduate from Mizoram University in 2023, holds a Bachelor of Technology (B.Tech) degree in the field of Electronics and Communication Engineering (ECE). With a strong academic foundation, he also obtained a diploma in Electronics and Tele Communication Engineering in 2020. Driven by a passion for advanced electronics and cutting-edge technology, Rakesh Kanjilal has embarked on a new academic journey. He is currently pursuing a Master's degree (M.Tech) at the prestigious National Institute of Technology, Sikkim, specializing in VLSI (Very-Large-Scale Integration) and Embedded Systems.

Professor **Reshmi Maity**[2] holds a faculty position in the Electronics and Communication Engineering department at Mizoram University.